\documentclass[preprint]{elsarticle} 
\usepackage{epsfig}
\usepackage{amssymb}

\def\eq#1{(\ref{#1})} 

\begin{document}

\title{Quantum tunneling radiation from self-dual black holes}

\author{C.A.S.Silva} 
\ead{calex@fisica.ufc.br}

\address{Instituto Federal de Educa\c{c}\~{a}o Ci\^{e}ncia e Tecnologia da Para\'{i}ba (IFPB),\\ Campus Campina Grande - Rua Tranquilino Coelho Lemos, 
671, Jardim Dinam\'{e}rica
I.
\\

} 
%


\begin{abstract}
We calculate the Hawking temperature for a self-dual black hole in the
context of quantum tunneling formalism.
\end{abstract}

\maketitle


%

\section{Introduction}

Black holes are putative objects whose gravitational fields are so strong that no physical bodies or signals can break free of their pull and
scape. In the seventies, through the Hawking demonstration that all black holes
emit blackbody radiation \cite{sw.hawking-cmp43}, the study of black holes obtained a  
position of significance going far beyond astrophysics, since, in the presence of a black hole strong gravitational field, the quantum nature of spacetime must be manifested.

One approach to quantum gravity, Loop Quantum
Gravity (LQG),  has given rise to models that afford
a description of the quantum spacetime revelled by a black hole. 
Acctually, string theory and
loop quantum gravity, lately, showed that the origin of
the black-hole thermodynamics must be related with the
quantum structure of the spacetime, bringing together the
developments in black-hole physics and the improvement
of our understanding on the nature of the spacetime in
quantum gravity regime

In particular, in loop quantum gravity context, a black hole metric, known as
the loop black hole (LBH), or self-dual black hole \cite{Modesto:2008im, Modesto:2009ve}, has the interesting property of self-duality that removes the black hole 
singularity and replaces it with
another asymptotically flat region. The issue of the thermodynamical of this kind of black hole has been investigated
in \cite{l.modesto-prd80, e.alesci-grqc-11015792, s.hossenfelder-grqc12020412}, and the dynamical aspects of the collapse 
and evaporation were studied in \cite{s.hossenfelder-prd81} where the habitual Hawking formalism to derive the black hole thermodynamical
properties was used.


By the way, since Hawking proved that black holes can radiate thermally \cite{sw.hawking-cmp43}, in a way that these objects are kinds of thermal system 
and have thermodynamic relations among the quantities describing them, several efforts in order to derive the temperature and entropy of black
holes have been done via various methods. While Hawking used the quantum field theory in curved spacetime in his original paper \cite{sw.hawking-cmp43},
there exist other methods which give the same predictions \cite{gthooft-npb256, sp.robinson-prl95, ec.vagenas-jhep0610, gw.gibbons-prd15}. Although all these methods have been successful in deriving the temperature or the 
entropy of certain types of black holes, it is not satisfactory in the sense that they do not reveal the dynamical nature of the radiation process,
since the background geometry is fixed mostly in these scenarios, including the Hawking's scenario.

In recent years, a semiclassical method has been
developed viewing Hawking radiation as a tunneling phenomena across the horizon \cite{mk.parikh-prl85, mk.parikh-hepth0402166, ec.vagenas-plb559}.
The essential idea is that the positive energy particle created just inside the
horizon can tunnel through the geometric barrier quantum mechanically, and it is observed as the Hawking flux at infinity.
The black hole tunneling method has a lot of strengths when compared to
other methods for calculating the temperature. To cite some of these strengths, we have that the tunneling method is a particularly interesting method for calculating
black hole temperature since it provides a dynamical model of the black hole
radiation. Besides, the calculations in this approach are straightforward and relatively simple, and the tunneling method 
is robust in the sense that it can be applied to a wide variety of exotic spacetimes \cite{qq.jiang-prd73, j.zang-plb638, r.kerner-prd73, 
l.zao-ctp47, m.agheben-jhep0505, sq.wu-jhep0603,
j.ren-cpl23, r.criscienzo-plb657, r.kerner-prd75, k.srinivasan-prd60, wg.unruh-prd14, mk.parikh-plb546, ajm.medved-prd66, s.shankaranarayanan-prd67}.

In this paper, we will use the tunneling formalism to investigate the thermodynamic properties of self-dual black holes.
%

%

\section{Self-dual black holes}

Loop quantum gravity is based on the formulation of classical general
relativity, which goes under the name of “new variables”, or “Ashtekar
variables”, that is in terms of an $su(2)$ 3-dimensional connection
A and a triad E. The basis states of LQG then are closed graphs the edges of which are
labeled by irreducible $su(2)$ representations and the vertices by $su(2)$ intertwiners.
One of the most significant result of loop quantum gravity is the
discovery that certain geometrical quantities, in particular area and volume, are represented by operators that have discrete eigenvalues.

The regular black hole metric that we will be using is derived from a simplified model of
LQG \cite{Modesto:2008im}. In this context, the quantum gravitational corrected metric provided by LQG is 

\begin{equation}
ds^{2} = - G(r)dt^{2} + F(r)^{-1}dr^{2} + H(r)d\Omega^{2} \label{self-dual-metric}
\end{equation}

\noindent with

\begin{equation}
 d\Omega^{2} = d\theta^{2} + \sin^{2}\theta\phi^{2}
\end{equation}

The metric functions are given by

\begin{equation}
G(r) = \frac{(r-r_{+})(r-r_{-})(r-r_{*})}{r^{4}+a_{0}^{2}} \; ,
\end{equation}

\begin{equation}
F(r) = \frac{(r-r_{+})(r-r_{-})r^{4}}{(r+r_{*})^{2}(r^{4}+a_{0}^{2})} \; ,
\end{equation}

\noindent and

\begin{equation}
 H(r) = r^{2} + \frac{a_{0}^{2}}{r^{2}}
\end{equation}

\noindent where

\begin{eqnarray*}
r_{+} = \frac{2Gm}{c^{2}} \;\; ; \;\; r_{-} = \frac{2Gm}{c^{2}}P^{2}
\end{eqnarray*}
\begin{eqnarray}
r_{*} = &\sqrt{r_{+}r_{-}} = 2mP
\end{eqnarray}

\begin{equation}
P = \frac{\sqrt{1+\epsilon^{2}} - 1}{\sqrt{1+\epsilon^{2}} +1} \;\; ; \;\; a_{0} = \frac{A_{\textrm{min}}}{8\pi} = \frac{\sqrt{3}}{2}\gamma\beta R_{P}^{2}
\end{equation}

In the next section, we will use the tunneling formalism to derive the Hawking temperature for a black hole described by the metric 
\eq{self-dual-metric}

\section{Tunneling from self-dual black holes}

The first black hole tunneling method developed was
the Null Geodesic Method used by Parikh and Wilczek\cite{mk.parikh-prl85, ec.vagenas-plb559}, which followed from
the work by Kraus and Wilczek \cite{p.kraus-gr-qc/9406042, p.kraus-npb433, p.kraus-npb437}. 
The other approach to black hole
tunneling is the Hamilton-Jacobi ansatz used by Angheben et al, which is an
extension of the complex path analysis of Padmanabhan et al 
\cite{k.srinivasan-prd60, s.shankaranarayanan-mpla16, s.shankaranarayanan-cqg19, t.padmanabhan-mpla19}.  
In this paper, we will work with this method, since it is more direct. Our calculations, using the Hamilton-Jacobi method, involves consideration of an emitted scalar particle, 
ignoring its self-gravitation, and assumes that its action satisfies the relativistic Hamilton-Jacobi equation.


To begin with, we have that, near the black hole horizon, the theory is dimensionally reduced to a $2$-dimensional theory \cite{Iso:2006ut,
Umetsu:2009ra} whose metric is just the $(t - r)$
sector of the original
metric while the angular part is red-shifted away. Consequently the near-horizon metric has the form

\begin{equation}
ds^{2} = - G(r)dt^{2} + F(r)^{-1}dr^{2} \label{r-metric} \; .
\end{equation}

\noindent Moreover, the effective
potential vanishes and there are no grey-body factors. However, the self-consistency of the approach can be seen by recalling that the emission 
spectrum obtained from
these modes
is purely thermal. This justifies ignoring the grey-body factors. 

Now, consider the Klein-Gordon equations

\begin{equation}
 \hslash^{2}g^{\mu\nu}\nabla_{\mu}\nabla_{\nu}\phi - m^{2}\phi = 0
\end{equation}

\noindent under the metric given by \eq{r-metric}

\begin{equation}
- \partial_{t}^{2}\phi + \Lambda \partial_{r}^{2}\phi + \frac{1}{2}\Lambda'\partial_{r}\phi - \frac{m^{2}}{\hslash^{2}}G\phi = 0
\end{equation}

\noindent where $\Lambda = F(r)G(r)$

Since the typical radiation wavelength is of the order of the black hole size, 
one might doubt whether a point particle description is appropriate. However, when
the outgoing wave is traced back towards the horizon, its wavelength, as measured by local
fiducial observers, is ever-increasingly blue-shifted. Near the horizon, the radial wavenumber
approaches infinity and the point particle, or WKB, approximation is justified \cite{mk.parikh-prl85}.

In this way, taking the standard WKB ansatz 

\begin{equation}
\phi(r, t)=e^{-\frac{i}{\hslash}S(r, t)}, \label{wkb}
\end{equation}

\noindent one can obtain the relativistic Hamilton-Jacobi equation with the limit $\hslash \rightarrow 0$,

\begin{equation}
(\partial_{t}S)^{2} - \Lambda (\partial_{r}S)^{2} - m^{2} = 0
\end{equation}

We seek a solution of the form

\begin{equation}
S(r,t) = -\omega t = W(r)
\end{equation}

Solving for W(r) yields

\begin{equation}
W = \int \frac{dr}{\Gamma} \sqrt{\omega^{2} - m^{2}G}
\end{equation}

\noindent where $\Gamma = \Lambda^{1/2}$

In this point, we will adopt the proper spatial distance,

\begin{equation}
d\sigma = \frac{dr^{2}}{\Gamma(r)}
\end{equation}

\noindent and, by taking the near horizon approximation

\begin{equation}
\Gamma(r) = \Gamma'(r_{H})(r-r_{H}) + ...
\end{equation}

\noindent we find that

\begin{equation}
\sigma = 2\frac{\sqrt{r - r_{H}}}{\Gamma'(r_{H})}
\end{equation}

\noindent where $0 < \sigma < \infty$.

In this way, the spatial part of the action function reads

\begin{eqnarray}
W &=& \frac{2}{\Gamma'(r_{H})}\int\frac{d\sigma}{\sigma}\sqrt{\omega^{2} - \frac{\sigma^{2}}{4}m^{2}G'(r_{H})\Gamma'(r_{H})} \nonumber \\
  &=& \frac{2\pi i \omega}{\Gamma'(r_{H})} + \textrm{real contribution}
\end{eqnarray}

In this way, the Hawing temperature for the self-dual black hole is given by

\begin{equation}
T_{H} = \frac{\omega}{ImS} = \frac{(2m)^{3}(1-P^{2})}{4\pi[(2m)^{4} +a_{0}^{2}]}
\end{equation}

\noindent The expression above for Black hole temperature has been found by \cite{l.modesto-prd80, e.alesci-grqc-11015792, s.hossenfelder-grqc12020412}. Moreover, in the limit of $m$ large it corresponds to the Hawking temperature, but goes to
zero  for $m \rightarrow 0$.

Moreover, the tunneling probability for a particle with energy $\omega$ is given by

\begin{equation}
\Gamma \backsimeq Exp[-2 Im S ] =  Exp\Big\{- \frac{\pi\omega[(2m)^{4} + a_{0}^{2}]}{m^{3}(1-P^{2})}\Big\}
\end{equation}

\noindent and the black hole entropy in this framework is given by \cite{l.modesto-prd80, e.alesci-grqc-11015792, s.hossenfelder-grqc12020412}

\begin{equation}
S = \frac{4\pi(1+P)^{2}}{(1-P^{2})}\Big[\frac{16m^{4} - a_{0}^{2}}{16m^{2}}\Big]
\end{equation}

\section{Conclusions}

In this work, we have used the Hamilton-Jacobi version of the tunneling formalism to derive the temperature and entropy of a self-dual black hole.
The result found out corresponds to that previously obtained by \cite{l.modesto-prd80, e.alesci-grqc-11015792, s.hossenfelder-grqc12020412}, where the usual Hawking calculation was applied. The expression found out to the
black hole temperature depends on the quantum of area $a_{0}$.

\expandafter\ifx\csname url\endcsname\relax \global\long\def\url#1{\texttt{#1}}
\fi \expandafter\ifx\csname urlprefix\endcsname\relax\global\long\def\urlprefix{URL }
\fi

\end{document}